\documentstyle[11pt,newpasp,twoside]{article}
\def\edcomment#1{\iffalse\marginpar{\raggedright\sl#1\/}\else\relax\fi}
\reversemarginpar

\begin{document}
\title{NGC1866: New analysis from HST-WFPC2 observations of the inner regions}

\author{Di Carlo$^1$ E., Raimondo$^{1,2}$ G., Brocato$^1$ E., Castellani$^3$ V., Walker$^4$ A.R.}

\affil{$^1$INAF-Osservatorio Astronomico di Collurania -- Teramo, Italy}
\affil{$^2$Dipartimento di Fisica, Universit\`a di Roma, La Sapienza, Roma, Italy}
\affil{$^3$INAF-Osservatorio Astronomico di Monte Porzio, Roma, Italy}
\affil{$^4$Cerro Tololo Inter-American Observatory, NOAO, La Serena, Chile}

\begin{abstract}
We report new results on the LMC globular cluster NGC 1866
obtained by analyzing F555W and F814W images from WFPC2@HST. On
the basis of the CMD we derive information on the cluster distance
and constraints on stellar evolution theory. Evidence of mass
segregation are found in the cluster core.
\end{abstract}

\section{Introduction}
The LMC cluster NGC 1866 is known to represent a workbench for stellar
evolution (Barmina et al. 2002). In particular, one may probe the
accuracy of stellar models in reproducing the evolution and structure
of intermediate mass stars. Recently, Barmina et al.  (2002) and Testa
et al. (1999) found opposite results in evaluating the efficiency of
the overshooting in the H-burning convective core in stars with mass
of the order of $4-5\,M_{\sun}$. Relying on our HST data we give new
indications and constraints on this open question.

\section{The new preliminary results from HST data}

Thanks to the high spatial resolution of HST, for the first time,
we obtained reliable photometry for stars in the very central
region of NGC 1866. Thus, we improve the statistics by almost
doubling the number of He-burning red giants (N$_{RG}$). The CMD
discloses a well defined main sequence (MS) down to $V\simeq 25$
mag. This made it possible to determine both the cluster distance
and reddening with the MS-fitting method. By adopting $Z=0.007$
and $Y=0.24$, we place NGC 1866 at $(m-M)_0 = 18.35$ with
$E(B-V)=0.06$ (Walker et al. 2001).

In comparing the whole CMD with a suitable set of isochrones, as
computed for this purpose assuming classical stellar models
(Schwarzschild criterion) and models with a moderate overshooting
($0.25 H_p$), we find in both cases a good agreement but with two
different age evaluations ($t \simeq 140 Myr$ and $200 Myr$,
respectively). Unexpectedly, the new observed luminosity function
(LF), normalized to N$_{RG} = 170$, sensitively differs (Fig. 1a)
from the LF given by Testa et al. (1999) and recently used by
Barmina et al. (2002).  The comparison of the observed LF with the
two theoretical LFs - as derived from our isochrones - suggests a
value smaller than $\sim 0.1 H_p$ for the adopted models (Fig.
1b). However, if binaries are taken into account (dotted line) the
classical models would better fit the data.

The above discussion refers to the LF obtained by considering all
the cluster, but an important result is found by studying the LF
in annular regions with selected distances from the cluster
center,
we find the evidence for a strong mass
segregation. Similar results are shown by de Grijs et al. (2002)
for other two young LMC clusters.

\section{Conclusions}
The HST data of NGC1866 provide new indications on the properties
of LMC young stellar systems. Here we outline preliminary results
on the CMD and on the LF. The complete discussion of the new
findings and the comparison with models will be given in a
forthcoming paper (Brocato et al. 2002, in preparation).

\begin{figure}[t]
\plotfiddle{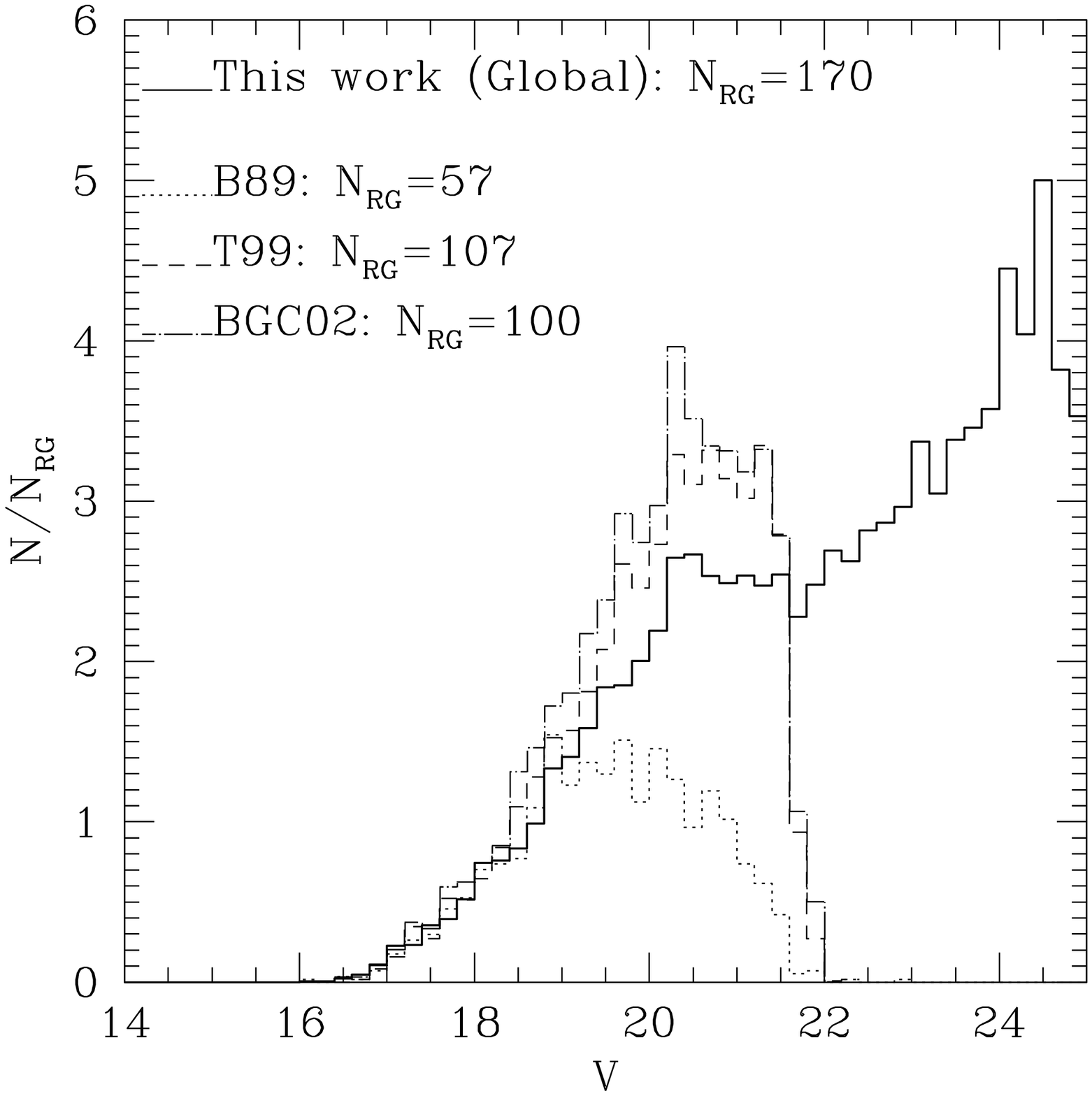}{0in}{0}{35}{35}{-220}{-230}
\plotfiddle{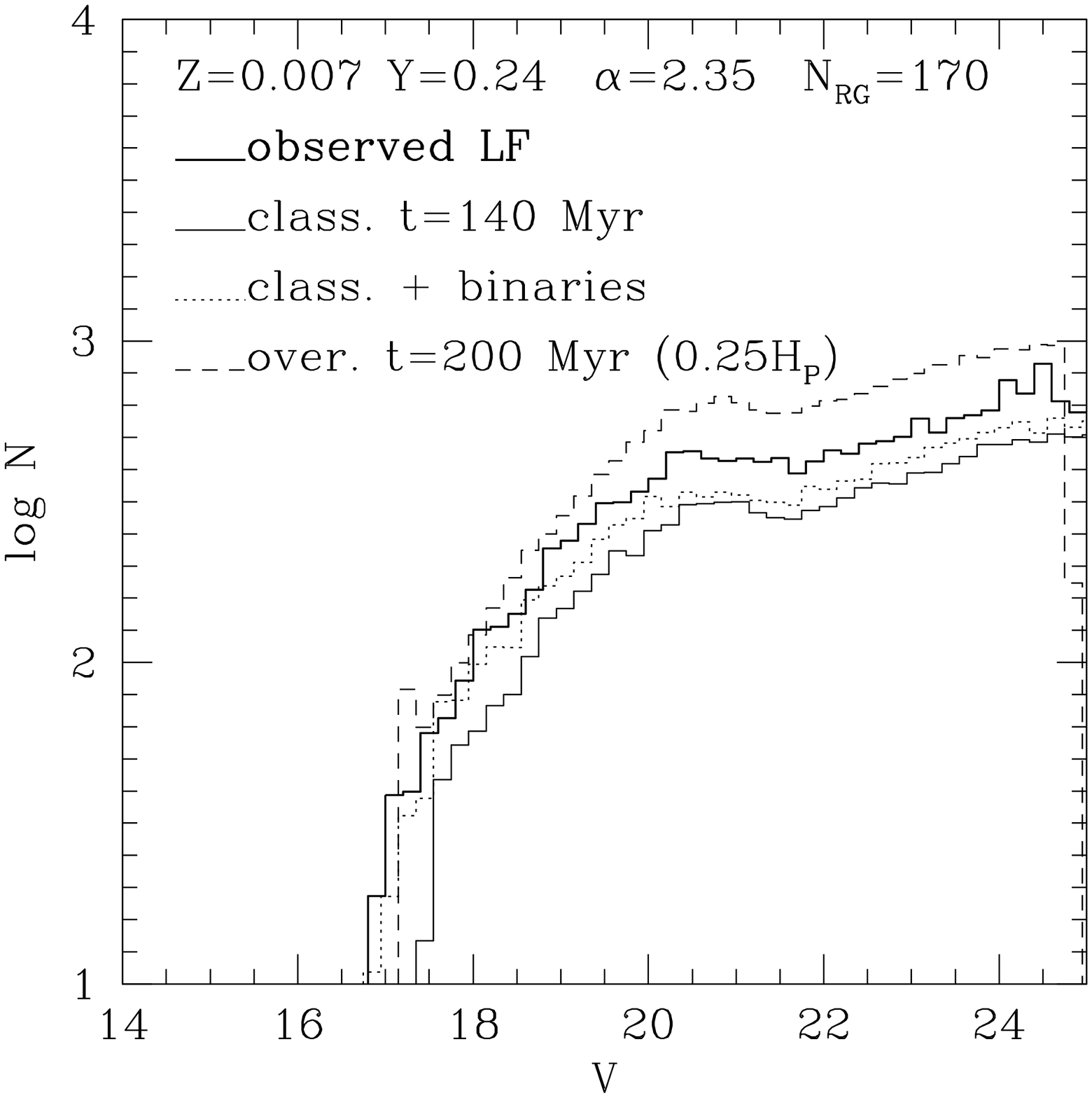}{0in}{0}{35}{35}{-10}{-205}
\vspace{5.5cm} \caption{\emph{Left panel:} the global normalized
LF is compared with selected LF adopted in previous works.
\emph{Right panel:} the global LF is compared to theoretical LFs
derived by fixing N$_{RG}=170$ (see text).  }
\end{figure}

\end{document}